\documentclass[amsmath,amssymb,aps,pra,twocolumn]{revtex4}
\usepackage{bm}
\usepackage{amsfonts}
\usepackage{amssymb}
\usepackage{graphicx}

\begin{document}

\title{Dynamical Casimir effect in oscillating media}
\author{Iwo Bialynicki-Birula}\email{birula@cft.edu.pl}
\affiliation{Center for Theoretical Physics, Polish Academy of Sciences\\
Al. Lotnik\'ow 32/46, 02-668 Warsaw, Poland}
\author{Zofia Bialynicka-Birula}\affiliation{Institute of Physics, Polish Academy of
Sciences\\
Al. Lotnik\'ow 32/46, 02-668 Warsaw, Poland}

\begin{abstract}
We show that oscillations of a homogeneous medium with constant material coefficients produce pairs of photons. Classical analysis of an oscillating medium reveals regions of parametric resonance where the electromagnetic waves are exponentially amplified. The quantum counterpart of parametric resonance is an exponentially growing number of photons in the same parameter regions. This process may be viewed as another manifestation of the dynamical Casimir effect. However, in contrast to the standard dynamical Casimir effect, photon production here takes place in the entire volume and is not due to time dependence of the boundary conditions or material constants.
\end{abstract}
\maketitle

\section{Introduction}

The results presented in this paper fall under the general category of dynamical Casimir effects. Hundreds of papers have been written on this subject --- a review by Dodonov \cite{dod} published seven years ago has 333 references (more recent references were added in \cite{dod1}). There is, however, a difference between all the numerous papers on this subject and our results. The standard notion of the dynamical Casimir effect involves regions with boundaries, as in the original static Casimir effect --- typical example is an oscillating mirror. In a few papers \cite{osa,y,js,abb,crm,cdlm,upss,pra} the case of changing properties of the medium was also considered. We show that an {\em infinite medium} in motion without any boundaries and with {\em time-independent} material constants may also lead to photon production. This effect is rather small, but it represents another manifestation of the dynamical Casimir effect.

The propagation of electromagnetic waves is affected by the motion of the medium. The simplest example of this effect has been predicted by Fresnel and observed by Fizeau in the XIX century (see a thorough review by Bolotovski\v{i} and Stolyarov \cite{bolot}). The Maxwell equations in a medium moving with velocity ${\bm v}$ retain their standard form:
\begin{subequations}\label{max}
\begin{align}
\partial_t{\bm D}({\bm r}, t)&=\nabla\!\times\!{\bm H}({\bm r}, t),\\
\nabla\!\cdot\!{\bm D}({\bm r}, t)&=0,\\
\partial_t{\bm B}({\bm r}, t)&=-\nabla\!\times\!{\bm E}({\bm r}, t),\\
\nabla\!\cdot\!{\bm B}({\bm r}, t)&=0,
\end{align}
\end{subequations}
but the constitutive relations between $({\bm D},{\bm B})$ and $({\bm E},{\bm H})$ are modified and read
\begin{subequations}\label{cr}
\begin{align}
{\bm D}+{\bm v}\!\times\!{\bm H}/c^2&={\epsilon}\left({\bm E}+{\bm v}\!\times\!{\bm B}\right),\\
{\bm B}-{\bm v}\!\times\!{\bm E}/c^2&={\mu}\left({\bm H}-{\bm v}\!\times\!{\bm D}\right).
\end{align}
\end{subequations}

We have assumed that the medium is homogeneous and isotropic, characterized in its rest frame by the material coefficients $\epsilon$ and $\mu$. The constitutive relations (\ref{cr}) were derived by Minkowski \cite{mink} and also by Einstein and Laub \cite{el}. Their derivations were based on Lorentz transformations and this might create an impression that they are valid only for a medium moving with a {\em constant velocity}. However, in the absence of dispersion these relations are valid for an arbitrary motion of the medium since they have a purely algebraic form (no derivatives). Therefore, they must be satisfied at {\em each space-time point} with the medium velocity taken at this point. This observation has already been made by Minkowski who stressed that his relations have a purely local character --- they hold at each time and at each point in space. In a recent book, Hehl and Obukhov \cite{ho} use geometric arguments to prove the validity of the ``Minkowski constitutive relations for arbitrarily moving matter.'' In the presence of dispersion the constitutive relations do not have the Minkowski form (\ref{cr}) because they connect field vectors at {\em different times}. Thus, our simple analysis cannot be applied in the frequency regions where the material constants depend significantly on the wave frequency. Therefore, our results hold only for frequencies that lie in dispersion-free regions.

In this paper we describe the influence of a harmonically oscillating dispersion-free medium on the propagation of electromagnetic waves. The motion of the medium leads to {\em time-dependent constitutive relations} which in turn result in parametric resonance. The existence of parametric resonance is a fairly general feature, but its influence on the propagation of electromagnetic waves depends on the detailed properties of the system. In our case, owing to the homogeneity of the medium, full analysis is possible and there is no need to resort to various simplifications (replacement of the electromagnetic field by a scalar field, restriction to one spatial dimension) that were introduced in most studies of the dynamical Casimir effect.

\section{Maxwell equations in a moving medium}

We shall seek the solution of Eqs.~(\ref{max}) and (\ref{cr}) in the form of the Riemann-Silberstein (RS) vector \cite{qed,pwf}
\begin{align}\label{f}
{\bm F}=\frac{\bm D}{\sqrt{2}\epsilon}+i\frac{\bm B}{\sqrt{2}\mu}.
\end{align}
The Maxwell equations take on the form:
\begin{align}
i\partial_t{\bm F}({\bm r}, t)&=\nabla\times{\bm G}({\bm r}, t),\label{max1}\\
\nabla\!\cdot\!{\bm F}({\bm r}, t)&=0,\label{max2}
\end{align}
where
\begin{align}\label{g}
{\bm G}=\frac{\bm E}{\sqrt{2}\mu}+i\frac{\bm H}{\sqrt{2}\epsilon}.
\end{align}
The Minkowski constitutive relations (\ref{cr}), expressed in terms of ${\bm F}$ and ${\bm G}$, are
\begin{align}\label{fg}
c{\bm F}+in{\bm v}\!\times\!{\bm F}=n{\bm G}+i\frac{\bm v}{c}\!\times\!{\bm G},
\end{align}
where $n=c\sqrt{\epsilon\mu}$ is the refractive index of the medium and ${\bm v}$ is the local velocity of the medium. The advantage of using the RS vector is the reduction of two equations to one equation without any loss of information --- we can always recover the original two equations by taking the real and imaginary parts. Equation (\ref{fg}) can be solved for ${\bm G}$ leading to constitutive relations in the form
\begin{align}\label{fg1}
{\bm G}=\frac{c}{n}\left({\bm F}+\frac{n^2-1}{c^2n^2-{\bm v}^2}{\bm v}\!\times\left({\bm v}\!\times\!{\bm F}+icn{\bm F}\right)\right).
\end{align}

The Maxwell equations (\ref{max1}) and (\ref{max2}) together with constitutive relations (\ref{fg1}) form a complete set of equations for the Riemann-Silberstein vector ${\bm F}$. The problem at hand is that of electromagnetic wave propagation in a medium with time-dependent constitutive relations. A distinctive feature of the present case is that the constitutive relations cannot be separated into electric and magnetic parts. After the separation of Eq.~(\ref{fg1}) into its real and imaginary parts, we see that the vector ${\bm E}$ (and similarly ${\bm H}$) depends on {\em both} ${\bm D}$ and ${\bm B}$.

From now on we will assume that the velocity of the medium depends only on time, and we express it in terms of a dimensionless vector ${\bm\beta(t)}$,
\begin{align}\label{vel}
{\bm v}(t)=c{\bm\beta(t)}.
\end{align}
In this case, the Maxwell equations (\ref{max1}) read
\begin{align}\label{max3}
&i\partial_t{\bm F}=-ic\gamma(t)[{\bm\beta(t)}\!\cdot\!\nabla]{\bm F}\nonumber\\&+(c/n)\alpha(t)\nabla\!\times\!{\bm F}-(c/n)\gamma(t){\bm\beta(t)}\!\times\!\nabla[{\bm\beta(t)}\!\cdot\!{\bm F}],
\end{align}
where
\begin{align}
\gamma(t)&=\frac{n^2-1}{n^2-\beta^2(t)},\label{alpha}\\
\alpha(t)&=1-\gamma(t)\beta^2(t).\label{gamma}
\end{align}

\section{Solutions of the Maxwell equations in a moving medium}

We shall construct the general solution of the Maxwell equations by performing the Fourier transformation with respect to the spatial variables. Since the coefficients in Eq.~(\ref{max3}) depend only on time, it is convenient to use the projections of the field on plane waves,
\begin{align}\label{ft}
{\bm f}({\bm k},t)=\int\!\frac{d^3r}{(2\pi)^{3/2}}e^{-i{\bm k}\cdot{\bm r}}{\bm F}({\bm r},t).
\end{align}
The components of the Fourier transform ${\bm f}({\bm k},t)$ obey the following set of ordinary differential equations:
\begin{align}\label{fteq}
\frac{d{\bm f}({\bm k},t)}{dt}&=-ic\gamma(t)[{\bm k}\!\cdot\!{\bm\beta}(t)]{\bm f}({\bm k},t)+\frac{c}{n}\alpha(t){\bm k}\!\times\!{\bm f}({\bm k},t)\nonumber\\
&+\frac{c}{n}\gamma(t){\bm k}\!\times\!{\bm\beta}(t)\,[{\bm\beta}(t)\!\cdot\!{\bm f}({\bm k},t)].
\end{align}
The first term on the right-hand side can be eliminated by the following change of the overall phase:
\begin{align}
{\bm f}({\bm k},t)=e^{-i\phi(t)}{\tilde{\bm f}}({\bm k},t),\label{phase}\\
\phi(t)=c\!\int_0^t\!dt'\,\gamma(t')[{\bm k}\!\cdot\!{\bm\beta}(t')]\label{phase1},
\end{align}
and we obtain
\begin{align}\label{fteq1}
\frac{d{\tilde{\bm f}({\bm k},t)}}{dt}\!=\!\frac{c}{n}\{\alpha(t){\bm k}\!\times\!{\tilde{\bm f}({\bm k},t)}
\!+\!\gamma(t){\bm k}\!\times\!{\bm\beta}(t)\,[{\bm\beta}(t)\!\cdot\!{\tilde{\bm f}({\bm k},t)}]\}.
\end{align}
Out of these three equations only two are independent since the amplitude ${\tilde{\bm f}({\bm k},t)}$ must obey the divergence condition ${\bm k}\cdot\!{\tilde{\bm f}({\bm k},t)}=0$. Therefore, we may always eliminate one equation from this set of three equations by taking the projections on two independent directions in the plane normal to ${\bm k}$.

In this paper, we assume that the velocity has a constant direction, specified by a unit vector ${\bm m}$, and only its length changes in time. Before discussing the general case we shall solve the evolution equations when the wave vector and the direction of motion are collinear.

\subsection{Propagation along the direction of velocity}

In this case, the evolution equation (\ref{fteq1}) reads
\begin{align}\label{fteq3}
d{\tilde{\bm f}({\bm k},t)}/dt=(c/n)\alpha(t){\bm k}\!\times\!{\tilde{\bm f}({\bm k},t)}.
\end{align}
It describes a precession of the vector ${\tilde{\bm f}({\bm k},t)}$ around the wave vector ${\bm k}$ with a varying angular velocity governed by the time-dependent velocity of the medium. One may check by inspection that the solution of Eq.~(\ref{fteq3}) is:
\begin{align}\label{solf0}
{\tilde{\bm f}}({\bm k},t)={\tilde{\bm f}}({\bm k},0)\cos\psi(t)+\frac{{\bm k}\!\times\!{\tilde{\bm f}}({\bm k},0)}{k}\sin\psi(t),
\end{align}
where
\begin{align}\label{psi}
\psi(t)=\frac{ck}{n}\int_0^t\!dt'\alpha(t'),
\end{align}
and $k=|{\bm k}|$. Since the length of the vector ${\tilde{\bm f}}({\bm k},t)$ does not depend on time, $|{\tilde{\bm f}}({\bm k},t)|^2=|{\tilde{\bm f}}({\bm k},0)|^2$, parametric resonance does not occur.

\subsection{Propagation along an arbitrary direction}

Unlike the simple solution obtained above in the special case, the solution in the general case cannot be given in terms of elementary functions. To reduce Eq.~(\ref{fteq1}) to a more amenable form, we expand the vector ${\tilde{\bm f}}({\bm k},t)$ in an orthonormal basis in the plane perpendicular to ${\bm k}$,
\begin{align}\label{fll}
{\tilde{\bm f}({\bm k},t)}={\bm n}_1({\bm k})f_1({\bm k},t)+{\bm n}_2({\bm k})f_2({\bm k},t).
\end{align}
We shall use the distinguished direction ${\bm m}$ to construct two unit vectors,
\begin{align}\label{ll}
{\bm n}_1({\bm k})=\frac{{\bm k}\!\times\!({\bm k}\!\times\!{\bm m})}{kk_\perp},\;\;\;{\bm n}_2({\bm k})=\frac{{\bm k}\!\times\!{\bm m}}{k_\perp},
\end{align}
where $k_\perp=|{\bm k}\times{\bm m}|$. These vectors, together with the unit vector ${\hat{\bm k}}$ in the ${\bm k}$ direction form an orthonormal basis,
\begin{align}\label{llk}
{\bm n}_1\times{\hat{\bm k}}={\bm n}_2,\;\;
{\hat{\bm k}}\times{\bm n}_2={\bm n}_1,\;\;
{\bm n}_2\times{\bm n}_1={\hat{\bm k}}.
\end{align}

The evolution equations (\ref{fteq1}) rewritten in terms of $f_1({\bm k},t)$ and $f_2({\bm k},t)$ read
\begin{subequations}\label{fteq2}
\begin{align}
\frac{df_1({\bm k},t)}{dt}&=\omega\alpha(t)f_2({\bm k},t),\\
\frac{df_2({\bm k},t)}{dt}&=-\omega\kappa(t)f_1({\bm k},t),
\end{align}
where
\begin{align}\label{kappa}
\kappa(t)=1-\cos^2\!\vartheta\,\gamma(t)\beta^2(t),
\end{align}
\end{subequations}
$\omega=ck/n$, and $\vartheta$ is the angle between the wave vector ${\bm k}$ and the direction of the velocity ${\bm m}$.

In Sec.~\ref{q} we will need a different decomposition of ${\tilde{\bm f}({\bm k},t)}$ that involves a complex polarization vector,
\begin{align}\label{fee}
{\tilde{\bm f}({\bm k},t)}={\bm e}({\bm k})f_+({\bm k},t)+{\bm e}^*({\bm k})f_-({\bm k},t),
\end{align}
where
\begin{subequations}\label{ef}
\begin{align}
{\bm e}({\bm k})&=\frac{\sigma{\bm n}_1-i\sigma^{-1}{\bm n}_2}{\sqrt{2}},\\
f_+({\bm k},t)&=\frac{\sigma^{-1}f_1({\bm k},t)+i\sigma f_2({\bm k},t)}{\sqrt{2}},\\
f_-({\bm k},t)&=\frac{\sigma^{-1}f_1({\bm k},t)-i\sigma f_2({\bm k},t)}{\sqrt{2}}.
\end{align}
\end{subequations}
At this moment we do not specify the value of the real parameter $\sigma$. The functions $f_\pm({\bm k},t)$ satisfy the following evolution equations:
\begin{subequations}\label{ftpm}
\begin{align}
\frac{df_+({\bm k},t)}{dt}&=-i\omega\left[\eta_+(t)f_+({\bm k},t)-\eta_-(t)f_-({\bm k},t)\right],\\
\frac{df_-({\bm k},t)}{dt}&=i\omega\left[\eta_+(t)f_-({\bm k},t)-\eta_-(t)f_+({\bm k},t)\right],
\end{align}
\end{subequations}
where
\begin{align}\label{eta1}
\eta_\pm(t)=\frac{1}{2}\left(\frac{\alpha(t)}{\sigma^2}\pm\kappa(t)\sigma^2\right).
\end{align}

To obtain concrete results we must specify the time dependence of the velocity.

\subsection{Motion with constant velocity}

Before analyzing accelerated motions, we shall solve the case of steady motion. This result will be later used as a reference point in the definition of the photon creation and annihilation operators when we will need the decomposition of the field operators into positive and negative frequency parts.

For a constant velocity, the functions $\alpha(t)$ and $\kappa(t)$ become time independent. If $\sigma=(\alpha/\kappa)^{1/4}$, then $\eta_-$ vanishes, and Eqs.~(\ref{ftpm}) decouple. Their solutions are
\begin{subequations}\label{solpm}
\begin{align}
f_+({\bm k},t)&=f_+({\bm k},0)e^{-i\omega_1t},\\
f_-({\bm k},t)&=f_-({\bm k},0)e^{i\omega_1t},
\end{align}
\end{subequations}
where $\omega_1=\omega\sqrt{\alpha\kappa}$.

For steady motion, the phase $\phi(t)$ is a linear function of time,
\begin{align}\label{lin}
\phi(t)=\omega_0t,\;\;\omega_0=ck\gamma\beta\cos\vartheta.
\end{align}
Thus, the complete solution of Eq.~(\ref{fteq}) is
\begin{align}\label{pnfp}
{\bm f}({\bm k},t)&=e^{-i\omega_0t}{\tilde{\bm f}}({\bm k},t),\\
{\tilde{\bm f}}({\bm k},t)&={\bm e}({\bm k})f_+({\bm k},0)e^{-i\omega_1t}+{\bm e}^*({\bm k})f_-({\bm k},0)e^{i\omega_1t}.
\end{align}
This formula does not always give a decomposition into positive and negative frequency parts because both $\omega_1+\omega_0$ and $\omega_1-\omega_0$ may change sign. However, this can happen only when the velocity $v$ exceeds the speed of light $c/n$ in the medium, because then $\omega_0$ may become larger than $\omega_1$. We shall restrict ourselves to moderate velocities when this interesting phenomenon does not occur.

\subsection{Harmonic oscillations}

We shall assume harmonic oscillations of the medium with the amplitude $A$ and the angular frequency $\Omega$,
\begin{align}\label{beta0}
\beta(t)=b\cos(\Omega t),
\end{align}
where $b=A\Omega/c$. The formulas for $\gamma(t)$ and $\alpha(t)$ take on the forms
\begin{align}
\gamma(t)&=\frac{n^2-1}{n^2-b^2\cos^2(\Omega t)},\label{gamma1}\\
\alpha(t)&=1-\gamma(t)b^2\cos^2(\Omega t).\label{alpha1}
\end{align}
For harmonic oscillations, the phase (\ref{phase1}) is
\begin{align}\label{phi}
\phi(t)=\frac{ck\cos\vartheta}{\Omega}
\frac{n^2-1}{\sqrt{n^2-b^2}}\arctan\left(\frac{b\,\sin(\Omega t)}{\sqrt{n^2-b^2}}\right).
\end{align}
It vanishes when the wave vector ${\bm k}$ is perpendicular to the direction of oscillations.

Equations (\ref{fteq2}) can, of course, be solved numerically. However, to get a better understanding of the general properties of the solutions, we shall reduce the two equations to a single equation for a parametrically driven mechanical harmonic oscillator. This will lead us, in the small-$b$ approximation, to the Mathieu equation whose properties can easily be explored with the use of MATHEMATICA. The appearance of the Mathieu equation in the dynamical Casimir effect has been noticed before \cite{jeong,dmm} in the case of an oscillating boundary.

\subsection{Mathieu equation}

To obtain the Mathieu equation, we make the substitutions
\begin{subequations}\label{subst}
\begin{align}
f_1({\bm k},t)&=\omega\sqrt{\alpha(t)}\,Q({\bm k},t),\\
f_2({\bm k},t)&=\frac{1}{\alpha(t)}\frac{d}{dt}\left[\sqrt{\alpha(t)}\,Q({\bm k},t)\right],
\end{align}
\end{subequations}
and we obtain the following equation for $Q({\bm k},t)$:
\begin{align}\label{mech}
\frac{d^2Q({\bm k},t)}{dt^2}+h({\bm k},t)Q({\bm k},t)=0,
\end{align}
where
\begin{align}
h({\bm k},t)
=\omega^2\alpha(t)\kappa(t)
-\sqrt{\alpha(t)}\,\frac{d^2}{dt^2}\!\left(\frac{1}{\sqrt{\alpha(t)}}\right).
\end{align}

After the decomposition of $h({\bm k},t)$ into the Fourier series, Eq.~(\ref{mech}) takes on the form of the Hill equation \cite{ince,ars,mag},
\begin{align}\label{hill}
\frac{d^2Q(\tau)}{d\tau^2}&+\left(\theta_0+2\sum_{l=1}^\infty\theta_l\cos(2l\tau)\right)Q(\tau)=0,
\end{align}
where $\tau=\Omega t$. We have omitted the argument ${\bm k}$ in $Q(\tau)$ but this dependence is still there through the parameters $\theta_l$. Under normal conditions the parameter $b$ is very small because the velocity of the medium is much smaller than $c$. We are, therefore, fully justified (see Fig.~\ref{fig2} for a graphic demonstration) to resort to the nonrelativistic approximation and keep only the first correction in the expansion of $h({\bm k},t)$ with respect to $b^2$. In this approximation Eq.~(\ref{mech}) becomes the Mathieu equation,
\begin{align}\label{mathieu}
\frac{d^2Q(\tau)}{d\tau^2}+\left[a-2q\cos(2\tau)\right]Q(\tau)=0,
\end{align}
whose properties are known in great detail \cite{as,gr,bp}. In our case the Mathieu parameters $a$ and $q$ have the values
\begin{subequations}\label{params}
\begin{align}
a&=\left(\frac{\omega}{\Omega}\right)^2\left(1-\frac{b^2}{2}\frac{n^2-1}{n^2}
\left(1+\cos^2\!\vartheta\right)\right),\\
q&=\left(\frac{\omega}{\Omega}\right)^2\frac{b^2}{4}\frac{n^2-1}{n^2}(1+\cos^2\!\vartheta)
-\frac{b^2}{2}\frac{n^2-1}{n^2}.
\end{align}
\end{subequations}

The most important property of the Mathieu equation is the appearance of alternating stable and unstable regions --- a clear signature of parametric resonance. In the unstable regions the solutions grow exponentially.

\section{Parametric resonance}

\begin{figure}
\centering
\vspace{0.5cm}

\includegraphics[scale=0.65]{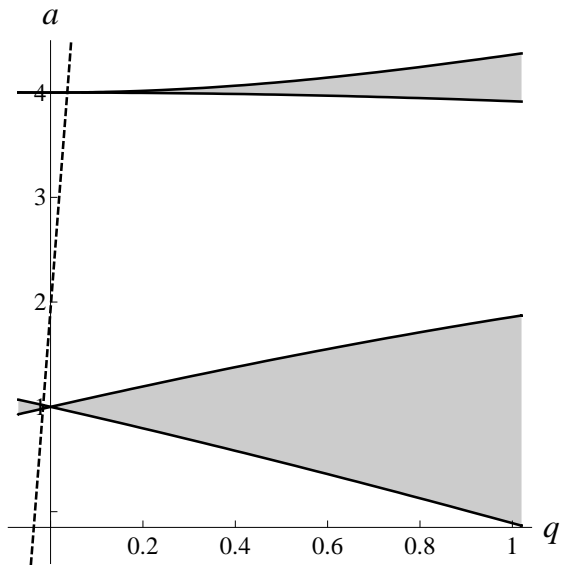}
\caption{Unstable regions (shaded areas) in the $(a,q)$ plane. The dashed line, defined by formula (\ref{sline}), corresponds to the values $n=2$, $b=0.3$, and $\vartheta=\pi/2$. The same values are chosen in Figs.~\ref{fig2}--\ref{fig4} and \ref{fig6}.}\label{fig1}
\end{figure}

\begin{figure}
\centering
\vspace{0.5cm}

\includegraphics[scale=1]{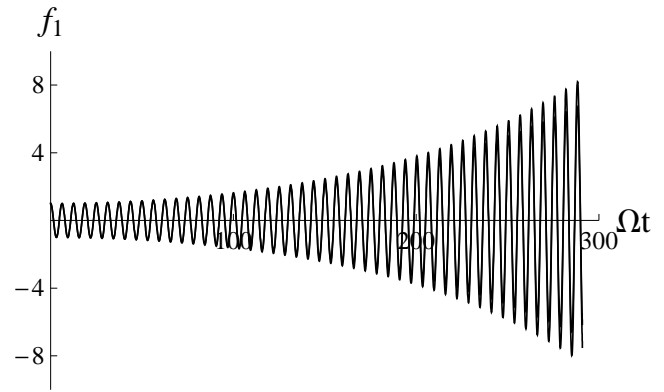}
\caption{In this plot we see two curves that lie almost exactly on top of each other. One represents the numerical solution of Eqs.~(\ref{fteq2}) and the other the analytic solution constructed from the corresponding Mathieu function for the same parameter values as in Fig.~\ref{fig1}. We have chosen the value $\omega/\Omega=1.016$ for which the imaginary part of the characteristic exponent is largest (see Fig.~\ref{fig4}). The exponential growth of the amplitude in time is clearly seen.}\label{fig2}
\end{figure}

The unstable regions in parametric resonance lie around the values of $\omega$ equal to multiples of $\Omega$. The location of these regions for the Mathieu equation is well known (see, for example, \cite{as}). The two lowest unstable regions in the plane of the parameters $a$ and $q$ are shown in Fig.~\ref{fig1}. A typical behavior of the amplitude in the region of parametric resonance is shown in Fig.~\ref{fig2}. The two almost coinciding curves are the best proof that the approximation of the solutions of Eqs.~(\ref{fteq2}) by the Mathieu functions is very good even for relatively large $b$.
To describe the properties of the electromagnetic field in a given medium we shall vary the wave parameters $\omega$ and $\vartheta$, keeping $n$ and $b$ fixed. Both $a$ and $q$ depend linearly on $\omega^2$. Therefore, all the points in the $(a,q)$ plane that differ only in the values of $\omega$ lie on a straight line defined by the equation
\begin{align}\label{sline}
&\frac{b^2}{4}\frac{n^2-1}{n^2}(1+\cos^2\!\vartheta)a\nonumber\\
&=\left(q+\frac{b^2}{2}\frac{n^2-1}{n^2}\right)
\left(1-\frac{b^2}{2}\frac{n^2-1}{n^2}\left(1+\cos^2\!\vartheta\right)\right),
\end{align}
as shown in Fig.~\ref{fig1}. Thus, parametric resonance occurs for those values of $a$ and $q$ where the dashed line in this figure crosses the shaded areas. With increasing $\omega$ the line crosses the consecutive unstable regions.

According to the Floquet theory (see, for example, \cite{as,ys0,ys}) every solution of the Hill equation (\ref{hill}) has the form
\begin{align}\label{floq}
Q(\tau)=e^{i\nu\tau}P(\tau)+e^{-i\nu\tau}P(-\tau),
\end{align}
where $P(\tau)$ is a periodic function with the period equal to $\pi$. The essential information about parametric resonance is contained in the characteristic exponent $\nu$. In the region of stable oscillations the characteristic exponent $\nu$ is real. The imaginary part of $\nu$ in an unstable region determines the rate of the exponential growth of solutions with time.

All numerical results presented in this paper are based directly on the solutions of Eqs.~(\ref{fteq2}) and (\ref{ftpm}). Their connection with the characteristic exponent appearing in the Floquet representation (\ref{floq}) is explained in the Appendix.

\begin{figure}
\centering
\vspace{0.5cm}

\includegraphics[scale=0.9]{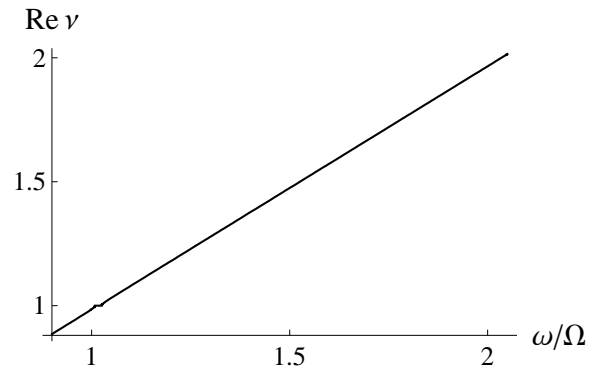}
\caption{Real part of the Mathieu exponent plotted as a function of $\omega/\Omega$ showing a tiny step at the first resonance (${\rm Re}\,\nu=1$). The step at the second resonance (${\rm Re}\,\nu=2$) is too small to be seen on this scale (see Fig.~\ref{fig4}).}\label{fig3}
\end{figure}

\begin{figure}
\centering
\vspace{0.5cm}

\includegraphics[scale=0.6]{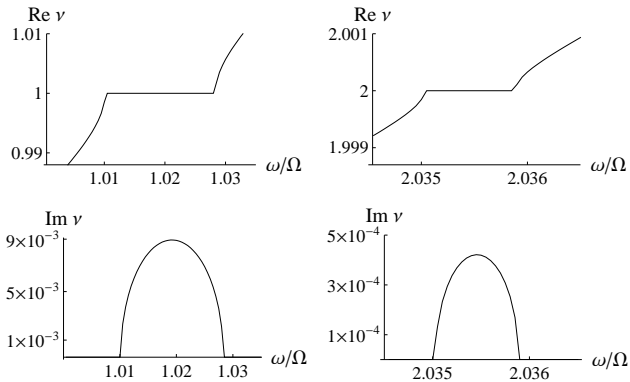}
\caption{Behavior of the real and imaginary parts of the characteristic exponent $\nu$ in two resonance regions (note the change of scale). The figures to the left correspond to the first resonance (${\rm Re}\,\nu=1$) and those to the right correspond to the second resonance (${\rm Re}\,\nu=2$).}\label{fig4}
\end{figure}

\begin{figure}
\centering
\vspace{0.5cm}
\includegraphics[scale=0.85]{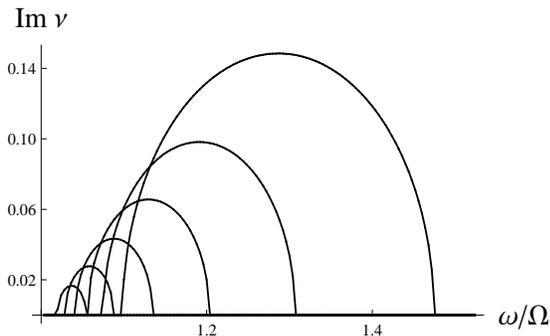}
\caption{Plots of the imaginary part of the characteristic exponent $\nu$ in the first resonance region for consecutive values of $b$ from 0.4 to 0.9.}\label{fig5}
\end{figure}

It is the general property of all Hill equations with real coefficients (see, for example, \cite{ars}) that in each unstable region the real part of $\nu$ is constant, equal to a natural number. Therefore, the plot of the real part of $\nu$ will exhibit characteristic steps, as seen in Figs.~\ref{fig3} and \ref{fig4}. Unfortunately, the consecutive regions of parametric resonance become narrower and narrower. Each time we move to the next higher resonance region the step shrinks by roughly a factor of $b^2$ and the imaginary part of $\nu$ diminishes at the same rate. Thus, the lowest resonance plays  the dominant role. For the lowest resonance, the values of the imaginary part of the characteristic exponent with varying $b$ are depicted in Fig.~\ref{fig5}. Their maximal values can be approximated by the formula ${\rm Im}\,\nu\approx 0.07 b^2+0.13 b^4$.

From the solutions $Q(t)$ of Eq.~(\ref{mech}) we may construct the Fourier transform of the RS vector,
\begin{align}\label{finq}
{\bm f}({\bm k},t)&=e^{-i\phi(t)}\Big({\bm n}_1({\bm k})\,\omega\sqrt{\alpha(t)}\,Q({\bm k},t)\nonumber\\
&+{\bm n}_2({\bm k})\frac{1}{\alpha(t)}\frac{d}{dt}\left[\sqrt{\alpha(t)}\,Q({\bm k},t)\right]\Big),
\end{align}
where, for clarity, we have reintroduced the dependence of $Q(t)$ on ${\bm k}$. The essential features are determined by the function $Q({\bm k},t)$, since the phase factor $\exp[-i\phi(t)]$ does not influence the discussion of parametric resonance.

The presence of parametric resonance for electromagnetic waves propagating in an oscillating medium leads to the amplification of those waves whose wave vectors lie in the resonance regions. The largest effect is for the waves with the frequencies close to $\Omega$ propagating in the direction perpendicular to the direction of oscillations. However, it would be difficult to achieve a significant amplification because it is governed by the exponential factor $\exp(0.1 b^2\Omega t)$. The essential parameter $b$ in this expression --- the maximal velocity of the medium divided by the speed of light --- is very small for realistic situations. It would take many periods of oscillations to achieve an observable amplification or significant photon production studied in the next section.

\section{Photon production}\label{q}

The dynamical Casimir effect may be viewed as an amplification of the vacuum fluctuations of the electromagnetic field that leads to the production of real photons. These effects are enhanced in the region of parametric resonance where the amplification grows exponentially with time.

The starting point of our analysis is the replacement of the classical field ${\bm F}({\bm r},t)$ by the field operator ${\hat{\bm F}}({\bm r},t)$ in the Heisenberg picture. Since this operator satisfies the same field equations as its classical counterpart, we can use the solutions of the field equations obtained in previous sections. The replacement of the classical functions by operators leads to the following representation of the field operator ${\hat{\bm F}}({\bm r},t)$
\begin{align}\label{repf}
{\hat{\bm F}}({\bm r},t)&=\int\frac{d^3k}{(2\pi)^{3/2}}e^{i{\bm k}\cdot{\bm r}}e^{-i\phi(t)}\nonumber\\
&\times\left[{\bm e}({\bm k}){\hat f}_+({\bm k},t)+{\bm e}^*({\bm k}){\hat f}_-({\bm k},t)\right],
\end{align}
where the polarization vector ${\bm e}({\bm k})$ is given by Eq.~(\ref{ef}). From the equal-time canonical commutation relations
\begin{align}\label{comrel}
\left[{\hat F}_i({\bm r},t),{\hat F}^\dagger_j({\bm r'},t)\right]
=-\frac{\hbar c}{n}\epsilon_{ijk}\partial_k\delta^{(3)}({\bm r}-{\bm r}'),
\end{align}
we obtain
\begin{subequations}\label{comrf}
\begin{align}
\left[{\hat f}_+({\bm k},t),{\hat f}^\dagger_+({\bm k}',t)\right]&=\hbar\omega\delta^{(3)}({\bm k}-{\bm k}'),\label{comrf1}\\
\left[{\hat f}_-({\bm k},t),{\hat f}^\dagger_-({\bm k}',t)\right]&=-\hbar\omega\delta^{(3)}({\bm k}-{\bm k}'),\label{comrf2}\\
\left[{\hat f}_-({\bm k},t),{\hat f}^\dagger_+({\bm k}',t)\right]&=0=\left[{\hat f}_-({\bm k},t),{\hat f}_+({\bm k}',t)\right].
\end{align}
\end{subequations}

\subsection{Creation and annihilation operators}

The opposite signs in the commutation relations (\ref{comrf1}) and (\ref{comrf2}) may suggest that ${\hat f}_+({\bm k},t)$ plays the role of an annihilation operator of one type of photon while ${\hat f}_-({\bm k},t)$ plays the role of a creation operator of another type of photon. However, such an interpretation requires separation into positive and negative frequency parts. This is possible only for a time-independent velocity. We can make use of this possibility if we assume that the velocity is constant at the beginning and at the end of the oscillatory motion, as shown in Fig.~\ref{fig6}a. In the regions of steady motion we can split the operator ${\hat{\bm f}}({\bm k},t)$ into positive and negative frequency parts. This enables us to define the incoming and outgoing creation and annihilation operators for both types of photon. We use again the value of $\sigma=(\alpha/\kappa)^{1/4}$ corresponding to the initial (and final) velocity,
\begin{subequations}\label{cran}
\begin{align}
t<0\nonumber\;\;\;{\hat f}_+({\bm k},t)&=\sqrt{\hbar\omega}\,{\hat a}_{\rm in}({\bm k})e^{-i\omega_1t},\\
{\hat f}_-({\bm k},t)&=\sqrt{\hbar\omega}\,{\hat b}_{\rm in}^\dagger(-{\bm k})e^{i\omega_1t}\\
t>t_n\nonumber\;\;\;{\hat f}_+({\bm k},t)&=\sqrt{\hbar\omega}\,{\hat a}_{\rm out}({\bm k})e^{-i\omega_1(t-t_N)},\\
{\hat f}_-({\bm k},t)&=\sqrt{\hbar\omega}\,{\hat b}_{\rm out}^\dagger(-{\bm k})e^{i\omega_1(t-t_N)},
\end{align}
\end{subequations}
where $t_N=2\pi N/\Omega$ specifies the number of periods of the oscillatory motion. The reversal of the sign of ${\bm k}$ in the second equation follows from the interpretation of ${\hat f}_-$ as the creation operator. The factor $\sqrt{\hbar\omega}$ has been extracted to secure the standard normalization of creation and annihilation operators. In the limiting case of a motionless medium we obtain the standard annihilation operators of photons with right and left helicity,
\begin{subequations}\label{lrh}
\begin{align}
{\hat a}_{\rm in}({\bm k})&={\hat a}_{\rm out}({\bm k})={\hat a}_R({\bm k}),\\
{\hat b}_{\rm in}({\bm k})&={\hat b}_{\rm out}({\bm k})={\hat a}_L({\bm k}).
\end{align}
\end{subequations}

The connection between the incoming and outgoing creation and annihilation operators can be obtained by solving the differential equations (\ref{ftpm}) for the operators ${\hat f}_\pm({\bm k},t)$. The time evolution from $t=0$ to $t=t_N$ leads to
\begin{subequations}\label{inout}
\begin{align}
&{\hat a}_{\rm out}({\bm k})=e^{-i\phi(t_N)}\nonumber\\
&\times\left[f^1_+({\bm k},t_N){\hat a}_{\rm in}({\bm k})+f^1_-({\bm k},t_N){\hat b}_{\rm in}^\dagger(-{\bm k})\right],\\
&{\hat b}_{\rm out}^\dagger(-{\bm k})=e^{-i\phi(t_N)}\nonumber\\
&\times\left[f^2_+({\bm k},t_N){\hat a}_{\rm in}({\bm k})+f^2_-({\bm k},t_N){\hat b}_{\rm in}^\dagger(-{\bm k})\right],
\end{align}
\end{subequations}
where the solutions $\{f^1_+({\bm k},t),f^1_-({\bm k},t)\}$ and $\{f^2_+({\bm k},t),f^2_-({\bm k},t)\}$ of Eqs.~(\ref{ftpm}) obey the following initial conditions:
\begin{subequations}\label{init}
\begin{align}
f^1_+({\bm k},0)=1,\;\;f^1_-({\bm k},0)=0,\\
f^2_+({\bm k},0)=0,\;\;f^2_-({\bm k},0)=1.
\end{align}
\end{subequations}
The mixing of the creation and annihilation operators (Bogoliubov transformation) is a clear signature of a two-mode squeezed vacuum state, accompanied by the creation of photon pairs. In order to calculate the photon production rate we assume that at $t=0$ there are no photons. It means that the state of system $|\psi\rangle$ is characterized by ${\hat a}_{\rm in}({\bm k})|\psi\rangle=0$ and ${\hat b}_{\rm in}({\bm k})|\psi\rangle=0$.

\subsection{Number of photons}

In infinite space the average number of produced photons is, of course, infinite. This infinity appears in the expression for the average number of outgoing photons, evaluated in the state $|\psi\rangle$, as the Dirac $\delta$ function with its argument equal to zero,
\begin{subequations}\label{avn}
\begin{align}
\langle\psi|{\hat a}_{\rm out}^\dagger({\bm k}){\hat a}_{\rm out}({\bm k})|\psi\rangle &=|f^1_-({\bm k},t_N)|^2\delta^{(3)}(0),\\
\langle\psi|{\hat b}_{\rm out}^\dagger({\bm k}){\hat b}_{\rm out}({\bm k})|\psi\rangle &=|f^2_+({\bm k},t_N)|^2\delta^{(3)}(0).
\end{align}
\end{subequations}
\begin{figure}
\centering
\vspace{0.7cm}
\includegraphics[scale=0.7]{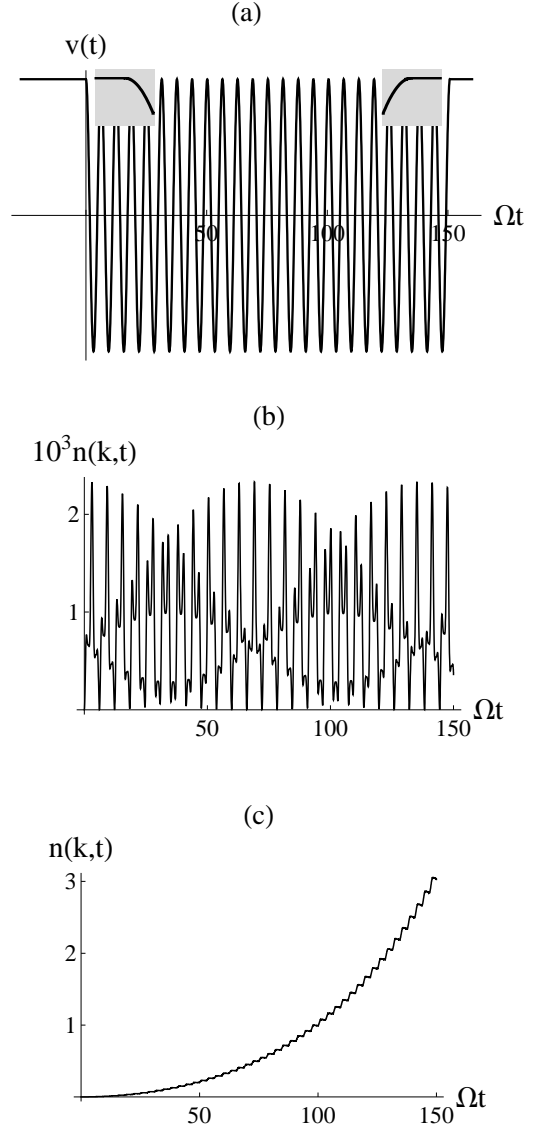}
\caption{(a) shows the general behavior of the velocity of the medium (in arbitrary units) as a function of $\Omega t$. The two insets show magnified views of areas where the constant velocity is smoothly joined to harmonic oscillations. (b) shows periodic oscillations of the photon number in the stable region ($\omega/\Omega=1.55$). (c) shows an exponential growth of the photon number in the region of parametric resonance ($\omega/\Omega=1.016$).}\label{fig6}
\end{figure}
To make use of these formulas we have to give a physical meaning of the dimensionless functions $|f^1_-({\bm k},t)|^2$ and $|f^2_+({\bm k},t)|^2$. First, let us note that owing to the symmetry of the evolution equations (\ref{ftpm}) under the transformation
\begin{subequations}\label{trans}
\begin{align}
f_+({\bm k},t)\to f^*_-({\bm k},t),\\
f_-({\bm k},t)\to f^*_+({\bm k},t),
\end{align}
\end{subequations}
the two functions in Eqs~(\ref{avn}) are equal (both types of photons are produced with the same probability). Their common value $n({\bm k},t)$,
\begin{align}\label{dens}
n({\bm k},t)=|f^1_-({\bm k},t)|^2=|f^2_+({\bm k},t)|^2,
\end{align}
is the number of photons of one type in the volume $h^3$ of the photon phase space. This can be shown with the help of the standard procedure of artificially restricting the volume to a finite cube of volume $V=L^3$ and imposing periodic boundary conditions. Then, $|f^1_-({\bm k},t)|^2$ and $|f^2_+({\bm k},t)|^2$ are the numbers of photons in the discrete modes specified by the wave vector ${\bm k}$ at time $t$. In the limit, when $L\to\infty$, we will find that $n({\bm k},t)$ is the density of photons in phase space. In other words, the total number of photons $n_R$ in a dispersion-free region $R$ of the phase-space is given by the formula
\begin{align}\label{nr}
n_R(t)=2\int_R\!\!\frac{d^3r\,d^3p}{h^3}n({\bm k},t),
\end{align}
where ${\bm p}=\hbar{\bm k}$.

Obviously, there is a big difference between the time dependence of $n({\bm k},t)$ in stable and unstable regions. In the stable regions we have periodic oscillations and in the unstable region we have an exponential growth (Fig.~\ref{fig6}).

\section{Conclusions}

We have demonstrated that, in addition to the two well-known mechanisms (time dependent boundary conditions or time dependent material coefficients) that lead to photon production through the dynamical Casimir effect, there exists a third mechanism. The main result of our study is the prediction of photon production in an oscillating medium without boundaries and with constant material coefficients. The density of photons in phase space given by Eq.~(\ref{dens}) grows exponentially in time in the regions of parametric resonance. These are the same regions where classical electromagnetic waves are amplified. Unlike the results given in most papers on the dynamical Casimir effect, our result is exact and based on full Maxwell equations in three dimensions.

\acknowledgments

We acknowledge the support by the Polish Ministry of Science and Higher Education under the Grant for the years 2008/2010.

\appendix
\section{Application of the Floquet theory to Eqs.~(\ref{ftpm})}

In order to obtain the characteristic exponents that identify the location of parametric resonance for a set of equations we need the monodromy matrix constructed from the fundamental solutions of these equations. Floquet theory states that the $2\times 2$ matrix constructed from the fundamental solutions defined in Eqs.~(\ref{init}) has the form
\begin{eqnarray}
\left(
\begin{array}{cc}f^1_+({\bm k},\tau)&f^1_-({\bm k},\tau)\\
f^2_+({\bm k},\tau)&f^2_-({\bm k},\tau)
\end{array}\right)={\hat X}(\tau)e^{{\hat K}\tau},
\end{eqnarray}
where ${\hat X}(\tau)$ is a periodic function of $\tau$ with period $\pi$ and ${\hat K}$ is a constant matrix. The fundamental solutions evaluated at $\tau=\pi$ form the monodromy matrix ${\hat M}={\hat X}(\pi)$ which determines all essential properties of every Floquet equation (see, for example, \cite{ys0,ys}). In particular, its eigenvalues $\mu_\pm$ determine the characteristic exponent $\nu$. The eigenvalue equation for ${\hat M}$ is
\begin{align}\label{eigen}
&\mu^2-\mu\left[f^1_+({\bm k},\pi)+f^2_-({\bm k},\pi)\right]/2\nonumber\\
&+\left[f^1_+({\bm k},\pi)f^2_-({\bm k},\pi)-f^1_-({\bm k},\pi)f^2_+({\bm k},\pi)\right]=0.
\end{align}
The last term, the determinant of the monodromy matrix, is the Wronskian of Eqs.~(\ref{ftpm}). It is equal to 1 --- its value at $\tau=0$ --- because the Wronskian is a constant of motion. Hence, the roots of Eq.~(\ref{eigen}) can be cast into the form
\begin{align}
\mu_\pm&=\left[f^1_+({\bm k},\pi)+f^2_-({\bm k},\pi)\right]/2\nonumber\\
&\pm i\sqrt{1-\left(f^1_+({\bm k},\pi)+f^2_-({\bm k},\pi)\right)^2/4}
=e^{\pm i\nu\pi},
\end{align}
where $\nu=\arccos\left[\left(f^1_+({\bm k},\pi)+f^2_-({\bm k},\pi)\right)/2\right]$ is, in general, complex.


\begin{thebibliography}{2}
\bibitem{dod} V. V. Dodonov, Adv. Chem. Phys. {\bf 119}, 309 (2001); arXiv:quant-ph/0106081.
\bibitem{dod1} V. V. Dodonov and A. V. Dodonov, J. of Russ. Laser Res. {\bf 26}, 445 (2005).
\bibitem{osa} Z. Bialynicka-Birula and I. Bialynicki-Birula, J. Opt. Soc. Am. B {\bf 4}, 1621 (1987).
\bibitem{y} E. Yablonovitch, Phys. Rev. Lett. {\bf 62}, 1742 (1989)
\bibitem{js} H. Johnston and S. Sarkar, Phys. Rev. A {\bf 51}, 4109 (1995).
\bibitem{abb} M. Artoni, A. Bulatov, and J. Birman, Phys. Rev. A {\bf 53}, 1031 (1996).
\bibitem{crm} M. Cirone, K. Rz{\c a}{\.z}ewski, and J. Mostowski, Phys. Rev. A {\bf 55}, 62 (1997).
\bibitem{cdlm} M. Crocce, D. A. R. Dalvit, F. C. Lombardo, and F. D. Mazzitelli, Phys. Rev. A {\bf 70}, 33811 (2004).
\bibitem{upss} M. Uhlmann, G. Plunien, R. Sch\"utzhold, and G. Soff, Phys. Rev. Lett. {\bf 93}, 193601 (2004).
\bibitem{pra} I. Bialynicki-Birula and Z. Bialynicka-Birula, Phys. Rev. A {\bf 77}, 052103 (2008).
\bibitem{bolot} B. M. Bolotovski\v{i} and S. N. Stolyarov, Sov. Phys. Usp. {\bf 17}, 875 (1975).
\bibitem{mink} H. Minkowski, Nachr. Ges. Wiss. G\"ottingen, math.-phys. Kl. {\bf 2}, 53 (1908).
\bibitem{el} A. Einstein and J. Laub, Ann. Phys. {\bf 26}, 532 (1908).
\bibitem{ho} F. W. Hehl and Yu. N. Obukhov, {\em Foundations of Classical Electrodynamics} (Birkh\"auser, Boston, 2003), p. 347.
\bibitem{qed} I. Bialynicki-Birula and Z. Bialynicka-Birula, {\em Quantum  Electrodynamics} (Pergamon, Oxford, 1975), p. 131.
\bibitem{pwf} I. Bialynicki-Birula, Photon wave function {\it Progress in Optics} Vol. 36, edited by E. Wolf (Elsevier, Amsterdam, 1996); arXiv:quant-ph/0508202.
\bibitem{jeong} J.-Y. Ji, H.-H. Jung, J.-W. Park, and K. S. Soh, Phys. Rev. A {\bf 56}, 4440 (1997).
\bibitem{dmm} D. A. R. Dalvit, F. D. Mazzitelli, and X. O. Millan, J. Phys. A: Math. Gen. {\bf 39} 6261 (2006).
\bibitem{ince} E. L. Ince, {\em Ordinary Differential Equations} (Dover, New York, 1956), p. 384.
\bibitem{ars} F. M. Arscott, {\em Periodic Differential Equations} (Pergamon, Oxford, 1964), p. 141.
\bibitem{mag} W. Magnus and S. Winkler, {\em Hill's Equation}, (Dover, New York, 1979).
\bibitem{as} {\em Handbook of Mathematical Functions}, edited by M. Abramowitz and I. A. Stegun (Dover, New York, 1964).
\bibitem{gr} I. S. Gradshteyn and I. M. Ryzhik, {\em Tables of Integrals, Series, and Products}, (Academic Press, New York, 2000).
\bibitem{bp} H. Bateman and A. Erdelyi, {\em Higher Transcedental Functions} (McGraw–Hill, New York, 1953), Vol. 3.
\bibitem{ys0} V. A. Yakubovitch and V. M. Starzhinski\v{i}, {\em Linear Differential Equations with Periodic Coefficients} (Wiley, New York, 1975).
\bibitem{ys} V. A. Yakubovitch and V. M. Starzhinski\v{i}, {\em Parametric Resonance in Linear Systems} (Nauka, Moscow, 1987).
\end{thebibliography}
\end{document}